# 2P-DNN : Privacy-Preserving Deep Neural Networks Based on Homomorphic Cryptosystem

Qiang Zhu[1][0000-0002-9293-0699] and Xixiang Lv [1]

[1] School of Cyber Engineering, Xidian University, Xian 710071, China
qzhu@stu.xidian.edu.cn
593304237@qq.com

**Abstract.** Machine Learning as a Service (MLaaS), such as Microsoft Azure, Amazon AWS, offers an effective DNN model to complete the machine learning task for small businesses and individuals who are restricted to the lacking data and computing power. However, here comes an issue that user privacy is exposed to the MLaaS server, since users need to upload their sensitive data to the MLaaS server. In order to preserve their privacy, users can encrypt their data before uploading it. This makes it difficult to run the DNN model because it is not designed for running in ciphertext domain.

In this paper, using the Paillier homomorphic cryptosystem we present a new Privacy-Preserving Deep Neural Network model that we called 2P-DNN. This model can fulfill the machine leaning task in ciphertext domain. By using 2P-DNN, MLaaS is able to provide a Privacy-Preserving machine learning service for users. We build our 2P-DNN model based on LeNet-5, and test it with the encrypted MNIST dataset. The classification accuracy is more than 97%, which is close to the accuracy of LeNet-5 running with the MNIST dataset and higher than that of other existing Privacy-Preserving machine learning models.

## 1 Introduction

Machine learning is an application of artificial intelligence (AI) that provides systems the ability to automatically learn and improve from experience without being explicitly programmed. Machine learning focuses on the development of computer programs that can access data and use it to learn for themselves. Machine learning is widely used in various fields and becomes an important part of AI.

Restricted to the lacking data and computing power, small businesses and individuals do not have the resources and ability of machine learning, especially in deep learning. To solve this problem, many cloud computing vendors provide a server called Machine Learning as a Service (MLaaS) such as Microsoft Azure Machine Learning Studio and Amazon AWS Machine Learning. With MLaaS, users can use the machine learning model and computing resource on the Cloud to fulfill their machine learning tasks.

However, the current MLaaS models do not consider to preserve user privacy, which prohibits its application and development. For example, intelligent medical on MLaaS may expose some private health information of patients. So, it is important to build a Privacy-Preserving MLaaS model which is able to run in the encrypted domain. Specifically, an MLaaS server runs the model with encrypted data and users get the predictions after decrypting the model output that is in ciphertext domain.

Aiming to preserve user privacy in MLaaS, we propose a Privacy-Preserving Deep Neural Network model (2P-DNN) based on the Paillier Homomorphic Cryptosystem. Our 2P-DNN model can run in the encrypted domain and give a prediction when input with encrypted data. The accuracy of our model is close to the same model in plaintext domain.

Our contributions are as follows:

1) We design a Privacy-Preserving Deep Neural Network model. Compared to the current Privacy-Preserving machine learning models[1,2,3,4], our model does not limit the input range and has the higher accuracy.

2) We build our 2P-DNN model based on LeNet-5, and test it with the encrypted MNIST dataset. The classification accuracy is not less than 99%, which is close to the accuracy of LeNet-5 running with the MNIST dataset and higher than other existing Privacy-Preserving machine learning models[4].

The rest of the paper is organized as follows. In Section 2, we review the related work. In Section 3, we introduce the Homomorphic Cryptosystem and the Deep Neural Network[5,6]. Then Section 4 presents our 2P-DNN model. In Section 5, we build our 2P-DNN model based on LeNet-5, and test it with the encrypted MNIST dataset. Finally, Section 6 concludes this paper.

## 2    Related Work

As for the importance of Privacy-Preserving machine learning model, there have been several excellent contributions in this field. These contributions mainly focus on two aspects, based on the traditional machine learning[1,2] and based on Deep Neural Network[3,4].

Some traditional machine learning models have linearity, which makes them able to run in the encrypted domain of Somewhat Homomorphic Encryption (SHE). By using this property, Raphael Bost *et al.* proposed a machine learning classification model over encrypted data[1]. They constructed three privacy-preserving classification algorithms: hyperplane decision, Naïve Bayes, and decision trees. They also combined these constructions with Adaboost. Louis J. M. Aslett *et al.* proposed a statistical machine learning model against encrypted samples[2]. By using Fully Homomorphic Encryption (FHE), they implemented a cryptographic random forest whose accuracy is lower than deep learning model.

In addition, some researchers use deep learning model to implement Privacy-Preserving MLaaS model. Nathan Dowlin *et al.* proposed CryptoNets [3]. They use the square function as the active function. But the square function results in that the computing resource cost becomes prohibitive as the number of layers increases, which is not adapted deep learning. To solve this problem, Florian Bourse *et al.* proposed a new framework (FHE-DiNN) for homomorphic evaluation of neural networks[4]. They refine the recent FHE algorithm proposed by Chillotti et al[7], in order to increase the message space and apply the sign function during the bootstrapping. However the input values of FHE-DiNNs are limited in {-1,1}, which greatly restricts its application.

## 3    Preliminaries

### 3.1    Paillier Homomorphic Cryptosystem

Homomorphic cryptosystem was first introduced by Rivest *et al.*[8], which allows one to directly operate the ciphertext and obtain the equivalent results in the plaintext domain without decrypting. The Paillier cryptosystem is a public key cryptosystem and a somewhat homomorphic cryptosystem[9]. The cryptosystem has been proved to be semantically secure. It provides homomorphic properties in terms of addictive and scalar multiplication, and is computationally comparable to RSA. The concrete Paillier cryptosystem is sketched as follows.

*Key Generation.* Let $p$, $q$ to be two large primes and $N = pq$. Let $Z_{N^2} = \{0, 1, \cdots, N^2 - 1\}$ and $Z_{N^2}^* \subset Z_{N^2}$ denotes the set including nonnegative integers which have multiplicative inverses module $N^2$. We also need to choose $g \in Z_{N^2}^*$ satisfying $\gcd\left(L\left(g^{\lambda} \bmod N^2\right), N\right) = 1$. Here, $\lambda = lcm(p-1, q-1)$ and we use it as the private key. The pair $(N, g)$ is the corresponding public key.

*Encryption.* Take a plaintext $m \in Z_N$, to encrypt m and get the ciphertext $c$, we compute:

$$c = E(m, r) = g^m r^N \bmod N^2 , \tag{1}$$

Where $r \in Z_N^*$ is an integer chosen randomly.

*Decryption.* When we get secret key $\lambda$, we can decrypt the ciphertext $c \in Z_{N^2}$ and get the plaintext $m$ as:

$$m = D(c, \lambda) = \frac{L\left(c^{\lambda} \bmod N^2\right)}{L\left(g^{\lambda} \bmod N^2\right)} \bmod N , \tag{2}$$

Where $L(u) = \frac{u-1}{N}$.

*Homomorphic Property.* The Paillier cryptosystem has addictive homomorphic property. That is,

$$c_1 \times c_2 \bmod N^2 = E\left(m_1, r_1\right) \times E\left(m_2, r_2\right)$$
$$= g^{\left(m_1+m_2\right)} \left(r_1 r_2\right)^N \bmod N^2 . \tag{3}$$

The exponential of $g$ is precisely the ciphertext of $m_1 + m_2$. In other words, we can get $m_1 + m_2$ after decrypting the result. Furthermore, we can get another from Equation (3) which is:

$$c_1 \times g^{m_2} \bmod N^2 = E\left(m_1, r_1\right) \times g^{m_2}$$
$$= g^{\left(m_1+m_2\right)} \left(r_1 r_2\right)^N \bmod N^2 . \tag{4}$$

We can also get $m_1 + m_2$ by decrypting this result.

In addition, the Paillier cryptosystem is homomorphic in terms of scalar multiplication. That is,

$$D\left[E\left[m, r\right]^k \bmod N^2\right] = k \cdot m \bmod N . \tag{5}$$

A modular exponential operation in ciphertext domain is equivalent to performing computing $k \cdot m$ in the plaintext.

We have to discuss the encryption on negative integers in this paper because the weight and bias of DNN might be negative. Paillier Cryptosystem does not involve negative integers, so some scholars design a novel method to solve this problem[10]. By dividing the encrypted domain into two equal spaces, one denotes positive integers and the other negative integers. The specific process can be described as:

*a)* Using this method to decrypt the data correctly needs to meet the condition in the formula.

$$2\sup\left(|m|\right) + 1 < N \tag{6}$$

where $sup(\cdot)$ denotes the biggest value in plaintext. $|m|$ is the absolute value of plaintext, $N$ is one of the Paillier public keys.

*b)* We compute this for encrypt a negative message $-m$:

$$E\left(-m\right) = E\left(-1 * m\right) = E\left(m\right)^{-1} = E\left(m\right)^{-1 \bmod N} \tag{7}$$

Where $m$ denotes a positive number in plaintext domain and $N$ is one of the Paillier public keys. We use the first half of $Z_N$ to denote the positive numbers and the second negative.

*c)* If the modulus of the cryptosystem is $N$, the $D[E[m]]$ is equal to $(m) \bmod N$ by $m'$. Let us denote $(m) mod N$ by $m'$. To decrypt the messages including negative numbers, one does as follows:

$$D_n(c) = \begin{cases} D(c) & (m < N/2) \\ D(c) - N & (m > N/2) \end{cases}.$$  (8)

    *d)*   Since the Paillier cryptosystem can only operate in the integer domain, we need to enlarge a non-integer value by multiplying it with an amplifying factor.

### 3.2 Deep Neural Networks

Deep Neural Network (DNN) is an important machine learning model in MLaaS. In recent years deep neural networks, such as Deep Auto Encoder, Convolutional Neural Network, and Recurrent Neural Networks, have shown to be capable of most machine learning tasks. Here, we briefly introduce the DNN structure and some key part of it.

  Fig.1 shows the DNN structure. We feed the input data to input layer. After a series of computation through the DNN, at the output layer we get the predicted result in terms of a task. At each layer, the output is sent to the next layer as its input, which is shown in Fig.2. We can write the output $y_i$ as:

$$y_j = f(\sum_i W_{ij} \times x_i + b).$$  (9)

Here, $x_i$ is the input, and $W_{ij}$ represents the weights. The function $f(\cdot)$ is the active function.

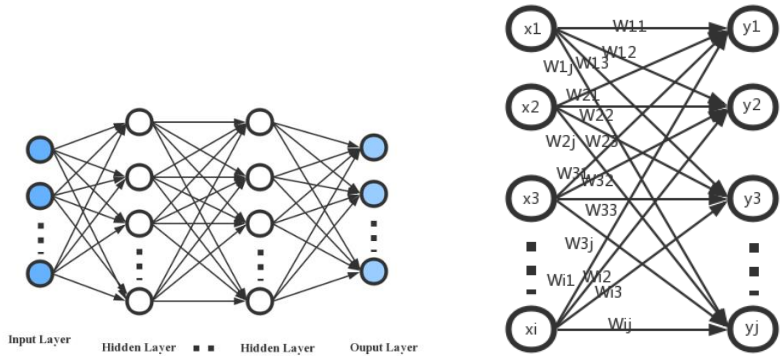

**Fig. 1.** DNN structure        **Fig. 2.** DNN linear transform

  The active function is a special part of DNN. It is used to describe the non-linear transform, whose performance is better than the linear transform. Rectified Linear Unit (ReLU) function is the most commonly used active function[11,12]. We can describe ReLU as $y = \max(0, x)$, which is shown in Fig.3.

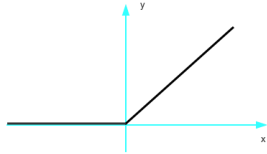

**Fig. 3.** ReLU function

CNN (Convolutional Neural Network) is widely used in Computer Vision and it is also an important model in MLaaS. CNN is a special kind of DNN, since it has a Convolution layer within each neuron. In the following, we only describe the special Convolution layers of CNN in some detail. A convolution layer accomplishes two operations, the convolution and the pooling. The Convolution can be formally described as follow:

$$a_{i,j} = \sum_m \sum_n w_{m,n} \times x_{i+m,j+n} + b ,\tag{10}$$

where $a_{i,j}$, $w_{m,n}$, $x_{i+m,j+n}$ and $b$ respectively represent the convolution output, the weights, the input and the bias. Pooling operation follows a convolution, and there are two pooling methods: max pooing and average pooling. The output of max pooling is the max value in pool. The average pooling is computing the average of values in pool.

## 4 Privacy-Preserving Deep Neural Networks Model

Considering the user privacy in MLaaS, we design 2P-DNN, a Privacy-Preserving Deep Learning model. In order to achieve an equivalent capability to the original DNN model, we do not change the DNN structure in the plaintext domain. In the 2P-DNN model, the input data is encrypted by users for protecting their privacy, and only users can decrypt and get the output. To fulfill the same MLaaS task as the plaintext domain, we need to improve most of the DNN algorithms, including linear transforms, active functions, convolution and pooling. The rest of this section focuses on the details of these improved algorithms.

### 4.1 Linear Transform

We use $E(\cdot)$ to represent the Paillier encryption algorithm. Thus, $E(x_i)$ is the encrypted input data, and $E(y_j)$ is the corresponding output that is also in ciphertext domain. According to the homomorphism of the Paillier algorithm, the linear transform of DNN algorithms can be improved as follow:

$$E[y_j] = E[\sum_i W_{ij} \times x_i + b]$$
$$= E[b] \times \prod_i E[x_i]^{W_{ij}} \quad . \tag{11}$$

Here, $W_{ij}$, $b$, $x_i$ and $y_j$ respectively represent the weights, bias, input and output. We can consider this computation as a linear transform layer. As shown in Equation (11), we can get the output $E(y_j)$ directly from $E(x_i)$ and $E(b)$ without decrypting. Then, the output $E(y_j)$ is taken as the input of the active function $f(\cdot)$.

## 4.2 Active Function

We choose ReLU as the active function of our model. Different from other active functions such as the Sigmoid function, ReLU is convenient to compute for our model. Since the ReLU function needs to judge the positivity or negativity of its input, we design an interactive protocol with which a server can compute ReLU by cooperating with its user. As minutely depicted in Fig. 4, the server-side sends the input data of ReLU to the client-side. Then, client-side decrypts it and sends the positivity or negativity of its plaintext to the server-side. The ReLU function of the server-side computes its output as $y = \begin{cases} x & x \geq 0 \\ 0 & x < 0 \end{cases}$.

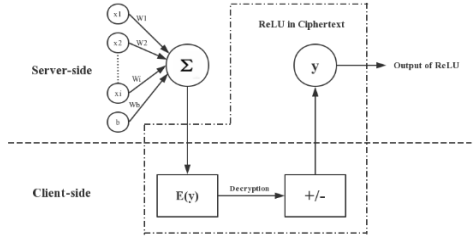

**Fig. 4.** Compute the ReLU function

## 4.3 Convolution and Pooling

Via convolution and pooling transform, CNN is able to achieve a superior performance. In terms of the homomorphism of the Paillier algorithm, the convolution algorithm in CNN can be improved as follow:

$$E(a_{i,j}) = E(\sum_m \sum_n w_{m,n} \times x_{i+m,j+n} + b)$$
$$= E[b] \times \prod_m \prod_n E[x_{i+m,j+n}]^{W_{m,n}} \quad , \tag{12}$$

Following a convolution, a pooling operation comes. There are two kinds of pooling, max pooling and average pooling. If we choose max pooling, we need to compare the values in ciphertext, which is difficult. In our model, we choose the average pooling as the pooling function, and design the pooling function in ciphertext as follow:

$$
\begin{aligned}
E[y] &= E[\frac{1}{s}\sum x_i] \\
&= \prod_i E[x_i]^{\frac{1}{s}}
\end{aligned}
, \tag{13}
$$

where $s$ is the stride of the pooling function.

## 5    2P-DNN Model on MNIST Dataset

In this section, we give the experimental evaluation with respect to the 2P-DNN model in the encrypted domain. We choose LeNet model in plaintext to fulfill the classification task on MNIST database which includes a great many handwritten digits. We implement our scheme using Python3.6 on a server with Intel® Xeon E5-2603 v4 CPU, 32G RAM, Nvidia Geforce GTX 1080Ti GPU.

### 5.1    Training the Model in Plaintext

In MLaaS, the server-side has a well trained model, with which MLaaS can provide classification service for users. Therefore, in this experimentation we first train a model with the MNIST dataset in plaintext. Then, we map this well trained model into a 2P-DNN model in ciphertext which is able to classify an encrypted object.

We use TensorFlow[13], an open source software library for DNN, to train a LeNet-5 model in the plaintext domain. As shown in Fig.5, this model consists of two convolution layers and two full connection layers. We use the MNIST[14] training dataset to train our MLaaS model based on LeNet-5. The MNIST training dataset has a training set of 60,000 examples. After training, the accuracy of our model achieves 99% on the MNIST testing dataset, which has a test set of 10,000 examples.

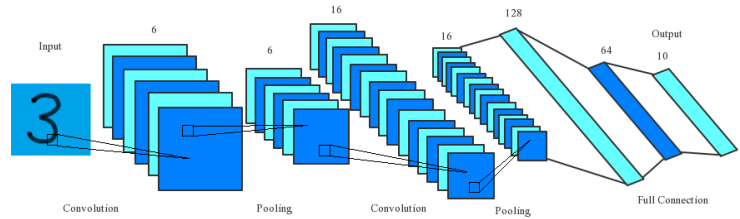

**Fig. 5.** LeNet-5 model

## 5.2    Encrypting Input Data Encryption

Aiming to verify the effectiveness of the proposed model, we use the MNIST database[15] of handwritten digits. This database includes a training set with 60,000 examples and a test set with 10,000 examples. Each image is size-normalized as 28x28 and a single pixel value is less than 256. In this experiment, we use the Paillier cryptosystem to encrypt an image at the pixel level. We encrypt each pixel by Paillier cryptosystem, and the algorithms come from Phe1.3.1[16], a library for Partially Homomorphic Encryption in Python. As demonstrated in Fig.6, the encrypted images do not reveal any trace of the handwritten digits. So, as we expect, this method is able to preserve user privacy.

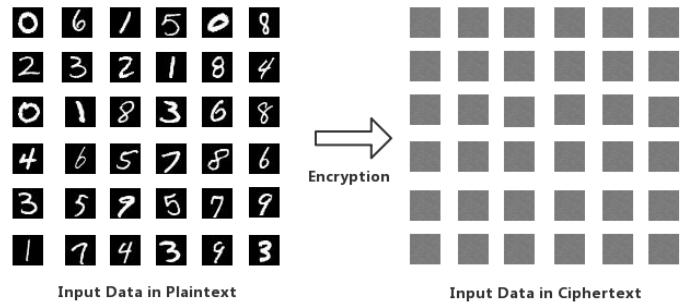

**Fig. 6.** Input data in plaintext and ciphertext

## 5.3    Implementing the Model in Ciphertext

Section 4 gives the main algorithms with which we can transform a fully trained LeNet-5 model into a 2P-DNN model. In this section, we present a privacy-preserving MLaaS based on these algorithms. As illustrated in Fig.7, this model is able to classify encrypted images.

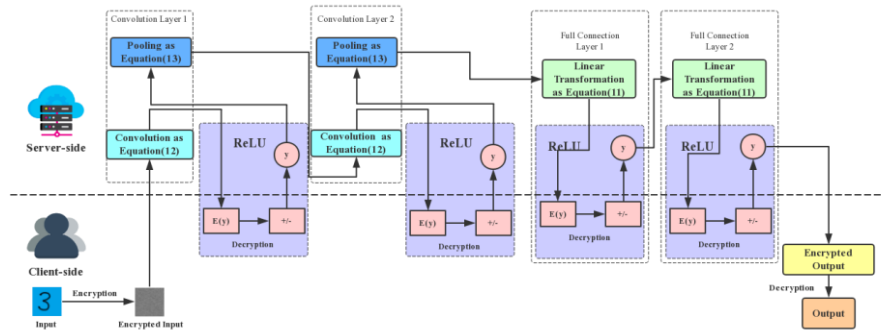

**Fig. 7.** Privacy-preserving MLaaS based on LeNet-5 in ciphertext

The details of our Privacy-preserving MLaaS is as follows:

1) The client-side encrypts a picture and initiates a request carrying this encrypted picture.

2) At the convolution layer, the input image is filtered as Equation (12), and then sent to the client-side. After receiving the return from the client-side, the server-side computes the ReLU output according to the plus or minus of the convolution output. After ReLU, the server-side does the pooling operation as Equation (13).

3) Following two rounds of convolution and pooling, it comes to the full connection layers. At each full connection layer, the server-side first computes the linear transformation, and then does the ReLU operation just like the convolution layers. The last full connection layer gives the predicted results in the encrypted domain.

4) At last, the server-side sends the encrypted result to the client-side, and the client-side can get the results in plaintext after decrypting it.

On the MNIST dataset encrypted by the Paillier cryptosystem, we get the accuracy of this model illustrated as the following table.

**Table 1.** Accuracy in plaintext and ciphertext.

| Convolution kernel size and number | The number of the full-connect layers | Accuracy in plaintext | Accuracy in ciphertext |
|---|---|---|---|
| (3*3, 32) | 512 | 99.0% | 98.6% |
| (3*3, 64) | 512 | 99.1% | 98.8% |
| (5*5, 32) | 512 | 99.0% | 98.7% |
| (5*5, 64) | 512 | 99.2% | 98.9% |
| (3*3, 32) | 1024 | 99.2% | 98.9% |
| (3*3, 64) | 1024 | 99.2% | 99.0% |
| (5*5, 32) | 1024 | 99.3% | 99.0% |
| (5*5, 64) | 1024 | 99.3% | 99.0% |

**Table 2.** Whether FHE-DiNNs and 2P-DNN can be applicate on classic datasets.

| Model | MNIST | CIFAR 10 | Iris Data Set | Wine Data Set |
|---|---|---|---|---|
| FHE-DiNNs | ✓ | ✗ | ✗ | ✗ |
| 2P-DNN | ✓ | ✓ | ✓ | ✓ |

As shown in Table 1, the accuracy of our model in ciphertext is close to that of LeNet-5 model in plaintext. The accuracy loss compared to plaintext model is due to the weight accuracy loss of the model during turning floats to integers. Despite this, our accuracy is a little higher than that of FHE-DiNNs model [7]. In addition, the inputs of FHE-DiNNs are limited in {-1,1}, it's application is greatly restricted. Most machine learning datasets are not applicable for FHE-DiNNs, since these datasets are not binary. Table 2 shows some classic datasets which can apply for the two models.

**Table 3.** Comparison of CryptoNets and 2P-DNN.

| Model | 2P-DNN | CryptoNets |
|---|---|---|
| Accuracy on MNIST | 99% | 99% |
| Active function | ReLU function | square function |
| Computational complexity of active function | O(1) | O(n) |

We also compare the CryptoNets and 2P-DNN in some aspects shown in Table 3. The accuracy of the two models are similar. However, the time cost of CryptoNets is more than that of 2P-DNN, because the active function of CryptoNets is square function and computing square function is slow in Ciphertext. The computing resource cost becomes prohibitive as the number of layers increases. Unlike the time cost of active function in CryptoNets, ours only depends on decryption operation, which is much faster than computation of the square function in ciphertext domain. Therefore, many DNN models are more suitable for our 2P-DNN model.

With respect to the efficiency, we use the Montgomery algorithm[17,18] to accelerate the modular exponentiation when images are encrypted. In terms of the structure, most of the signal processing model in the encrypted domain cannot be implemented by parallel computing, because each step of the algorithms is attached to the others[19,20]. However, our model can run by parallel computing, because the single neure in one layer is independent. The speed of our model can increases by more than 80% with these acceleration algorithms.

## 6  Conclusion

In this paper, we present a Privacy-Preserving Deep Neural Networks Based on Homomorphic Cryptosystem. This model can fulfill the machine learning task in ciphertext domain. So the MLaaS can use this model to protect the user privacy. The experiments demonstrate that our model can reach the same accuracy compared to the model in the plaintext. And the time cost of our model is much lower than other signal processing models in the encrypted domain.